\documentclass[conference]{IEEEtran}
\IEEEoverridecommandlockouts
\usepackage{cite}
\usepackage{amsmath,amssymb,amsfonts}
\usepackage{algorithmic}
\usepackage{graphicx}
\usepackage{textcomp}
\usepackage{xcolor}
\usepackage{bbm}
\usepackage{siunitx}
\usepackage{amsmath}
\usepackage{booktabs}
\usepackage{bm}
\usepackage{amsfonts}
\usepackage{subcaption}
\usepackage{balance}
\usepackage{float}
\usepackage{pgfplots}%
    \usepgfplotslibrary{units} 
    \pgfplotsset{compat=1.9}
\makeatletter
\pgfplotsset{
unit code/.code 2 args=
    \begingroup
    \protected@edef\x{\endgroup\si{#2}}\x
}

\def\BibTeX{{\rm B\kern-.05em{\sc i\kern-.025em b}\kern-.08em
    T\kern-.1667em\lower.7ex\hbox{E}\kern-.125emX}}

\newcommand{\LSD}{\operatorname{LSD}}

\begin{document}

\title{Towards HRTF Personalization using Denoising Diffusion Models}

\author{\IEEEauthorblockN{Juan Camilo Albarracín Sánchez, Luca Comanducci, Mirco Pezzoli, Fabio Antonacci} \IEEEauthorblockA{
Dipartimento di Elettronica, Informazione e Bioingegneria (DEIB), Politecnico di Milano\\
Piazza Leonardo Da Vinci 32, 20133 Milan, Italy\\
Email: juancamilo.albarracin@mail.polimi.it, luca.comanducci@polimi.it, mirco.pezzoli@polimi.it, fabio.antonacci@polimi.it
}\thanks{This work has been funded by ``REPERTORIUM project. Grant agreement number 101095065. Horizon Europe. Cluster II. Culture, Creativity and Inclusive Society. Call HORIZON-CL2-2022-HERITAGE-01-02.''
This work was partially supported by the European Union under the Italian National Recovery and Resilience Plan (NRRP) of NextGenerationEU, partnership on ``Telecommunications of the Future'' (PE00000001 - program ``RESTART'')}}

\maketitle

\begin{abstract}
Head-Related Transfer Functions (HRTFs) have fundamental applications for realistic rendering in immersive audio scenarios. However, they are strongly subject-dependent as they vary considerably depending on the shape of the ears, head and torso. 
Thus, \textit{personalization} procedures are required for accurate binaural rendering. 
Recently, Denoising Diffusion Probabilistic Models (DDPMs), a class of generative learning techniques, have been applied to solve a variety of signal processing-related problems. 
In this paper, we propose a first approach for using DDPM conditioned on anthropometric measurements to generate personalized Head-Related Impulse Response (HRIR), the time-domain representation of HRTF. 
The results show the feasibility of DDPMs for HRTF personalization obtaining performance in line with state-of-the-art models.
\end{abstract}


\section{Introduction}
Personalization of Head-Related Transfer Functions (HRTFs) is a key component in the domain of virtual auditory displays, which is extensively used in current development areas, such as Virtual Reality (VR), Augmented Reality (AR), gaming and auditory research~\cite{iidaheadrelated2019}. 
HRTFs, or their time-domain counterpart denoted as the Head-Related Impulse Responses (HRIRs), are crucial for creating immersive and realistic audio experiences as they are descriptors of how a person receives sound from a radiating source in space and they are highly sensitive to the listener's physical characteristics \cite{CA_book}. 
Thus HRTFs are unique from person to person. 
Obtaining accurate individual HRTFs is possible through acoustic measurements, however, the procedure involves specialized equipment and dedicated spaces, such as anechoic chambers, making the process not only time-consuming, but also expensive and inaccessible for the general consumer. 
Many applications rely on the use of generic HRTFs to render spatial audio, but it has been proven that the outcome of this kind of system results in a poor auditory experience, as localization precision is low, there is front-back confusion, and inside-head sound effect \cite{moller_fundbinaural,Blauert_spatialhearing}.

The complexity and high cost associated with HRTF acquisition highlights the need for efficient computational methods to model them. 
Numerical simulation methods have been raised as possible solutions to model HRTFs from scanned meshes \cite{huttunen2014rapid}, but these methods have the drawback of coming with large computational complexity and they take several hours to calculate a full individual set of HRTFs, making them unsuitable for low-resource devices \cite{zhang2021dnnbem}. 
As an alternative approach, \cite{zotkin_hrtf_2003} investigates the matching of specific anthropometric ear parameters with an HRTF dataset using the minimum distance criterion to customize individual HRTFs, making this method much more computationally efficient. 
However, existing HRTF datasets are not sufficiently large, and the HRTFs generated using anthropometric parameter matching are merely approximations. 
Consequently, they often lack the accuracy required for precise individualized spatial audio reproduction. 

Recently, machine learning has demonstrated relevant results in several audio and acoustics related tasks \cite{olivieri2021audio,cobos2022overview, koyama2024physics}, including innovative approaches to tackle HRTF personalization~\cite{chen_autoencoding_2019, chun_deep_2017,zhi2022towards,hogg2024hrtf,masuyama2024niirf,picinali2022sonicom}. 
Most of these works rely on Deep Neural Networks (DNNs) to link the listener's morphology to HRTF features, offering a more sophisticated and potentially accurate method of personalization \cite{chen_autoencoding_2019, chun_deep_2017,zhi2022towards,hogg2024hrtf}. 
By leveraging the existing datasets and the ability to model complex, non-linear relationships, these works provided promising results for improving the precision of HRTF customization. 
State-of-the-art works explored the feasibility of global HRTF prediction in the magnitude domain, using Spherical Harmonics (SH) as a compact representation method and anthropometric measurements as inputs \cite{wang2020global, wang2022predictingglobalheadrelatedtransfer}. 
Thus, deep learning-based models are designed to map anthropometric measurements to the SH representation of the HRTF.

In this paper, we explore the application of denoising diffusion probabilistic models (DDPMs) for HRIR personalization. 
Denoising probabilistic models are a class of generative models known for creating high-quality samples from noise \cite{ho2020denoising}. 
Recently, besides its demonstrated image generation capabilities, DDPMs have sprouted as an alternative approach for different sound synthesis problems and acoustic relevant scenarios \cite{moliner2023solving}, including dereverberation \cite{moliner2024buddy}, timbre transfer \cite{comanducci2024timbre}, vocoders \cite{kong2021diffwave} and sound field reconstruction \cite{miotello2024reconstruction}. 
By leveraging these models, we aim to generate personalized HRIRs, along with the sound source spatial information and the anthropometric measurements derived from an existing database \cite{HUTUBS_dataset_2019}.

The primary contributions of this work are threefold: (1) We introduced a method for HRIR personalization using denoising diffusion models that, to the best of our knowledge, there has been little or no work performing HRIR personalization using this particular model; (2) We evaluate the efficacy of our approach employing objective measurements comparing the generated HRIRs with ground truth data in both time and frequency domains and we contrast state-of-the-art investigations; (3) We identify and discuss the limitations and potential improvements for the future research in this domain.

\section{Problem formulation}
\begin{figure*}[htb]
    \centering
    \begin{subfigure}[b]{0.45\linewidth}
        \includegraphics[width=\textwidth]{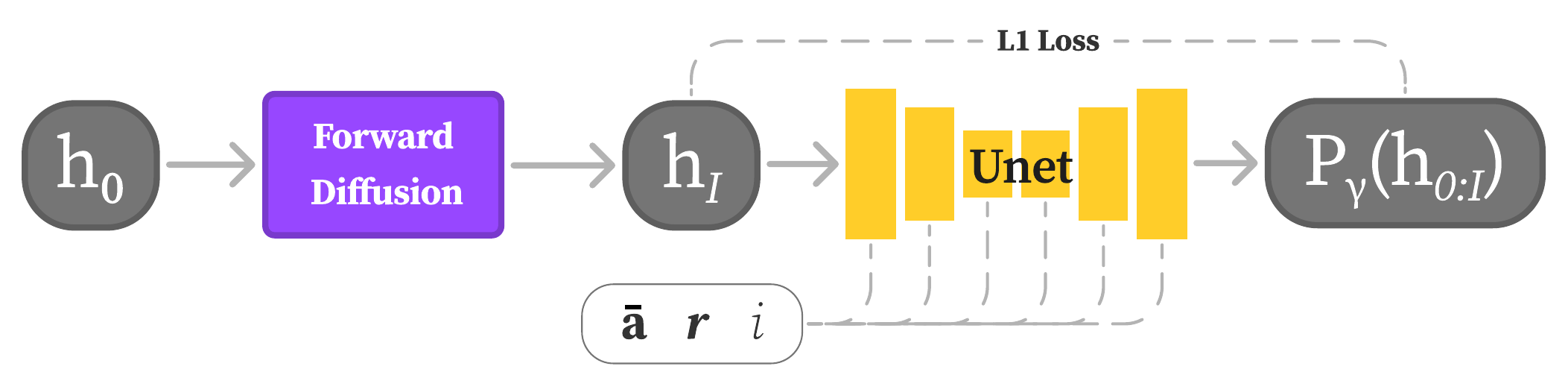}
        \caption*{(a)}
    \end{subfigure}
    \hspace{0.05\linewidth}
    \begin{subfigure}[b]{0.45\linewidth}
        \includegraphics[width=\textwidth]{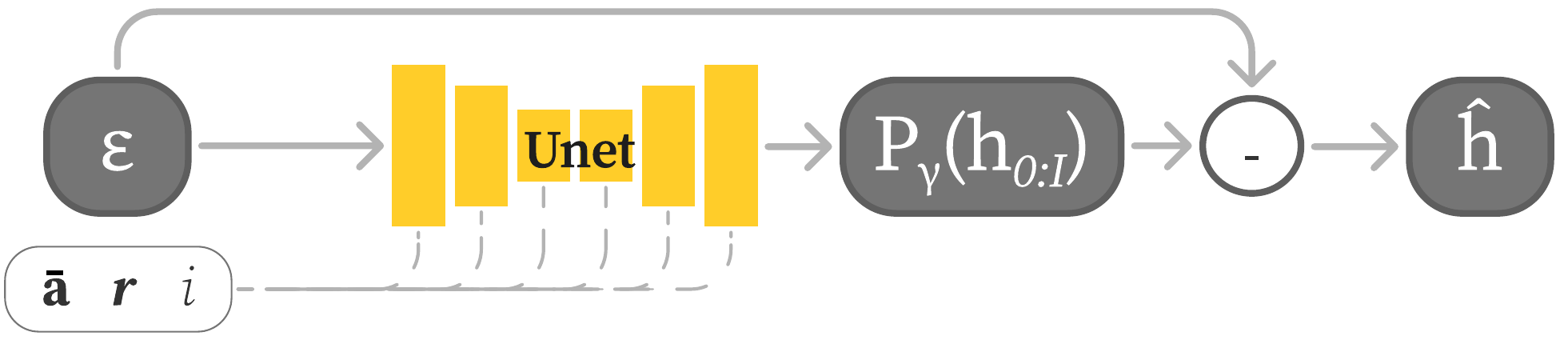}
        \caption*{(b)}
    \end{subfigure}
    \caption{Outline of the (a) training and (b) inference stages of the proposed method for HRIR personalization. Note how the conditioning information is embedded at each encoding/decoding block.}
    \label{fig:architecture}
\end{figure*}
\subsection{Signal Model}
Let us consider a sound source emitting in the far field from a listener, the pressure measured at one of the eardrums of the subject is represented by the so-called Head-Related Impulse Response (HRIR) defined as \cite{hongmei_vas_2008}
\begin{equation}
    p (t, \bm{r}, \mathbf{a})= \int_{\mathbbm{R}}h(t-\tau, \bm{r}, \mathbf{a})p_0(\tau)d\tau,
\label{eq:hrir_model}
\end{equation}
where $\bm{r}=(\theta, \phi)$ represents the source direction of arrival (DOA) with azimuth $\theta\in[0,2\pi)$ and elevation $\phi\in[-\pi/2, \pi/2]$, $p_0$ is the pressure relative to the free field at the position of the eardrum, and  $\mathbf{a}\in \mathbb{R}^{N\times 1}$ are the anthropometric features of the listener. 
As a matter of fact, HRIRs are individualized to each listener according to the physical characteristics of the subject \cite{Blauert_spatialhearing}. 
Such characteristics can be typically represented by anthropometric features $\mathbf{a}$ which contains $N$ features concerning, for example, the ear shape, head size, and torso dimensions. 
It is worth to note that the well-known HRTF is defined as the Fourier-transformed HRIR \eqref{eq:hrir_model}, making them a Fourier pair.

\subsection{Problem Definition}
The goal of the proposed model is to estimate the personalized HRIR \eqref{eq:hrir_model} given the anthropometric information of the desired subject and the target DOA. 

Based on \eqref{eq:hrir_model}, $h(\cdot)$ has to be reconstructed in order to accurately model the HRIR for a given anthropometric configuration $\mathbf{a}$. 
In practice, the proposed task can be formulated as an inverse problem, whose solution is found as
\begin{equation}
\begin{aligned}
    \mathrm{\bm{\gamma}^*} &= \arg \min_{\bm{\gamma}} J(\bm{\gamma})  = \\
    &E\left( f_{\bm{\gamma}}\left(t, \bm{r},\mathbf{a} \right),  h(t,\bm{r},\mathbf{a})\}\right)_{(\bm{r}, \mathbf{a}) \in \mathcal{D}} , 
\end{aligned}\label{eq:problem_def}
\end{equation}
where $\mathcal{D}$ denotes the set of available DOAs and anthropometric features. 

The term $f_{\bm{\gamma}}\left(t, \bm{r},\mathbf{a} \right)$ represents a function whose learnable parameters $\bm{\gamma}$ are optimized to estimate the HRIRs based on the available measurements. 
In fact, in \eqref{eq:problem_def}, $E(\cdot)$ is a data fidelity term, e.g.,  the mean absolute error (MAE), which quantifies the difference between the reconstructed and observed HRIR. 
While the optimization is performed in order to minimize the error with respect to the available data $h(t, \bm{r}, \mathbf{a})_{(\bm{r}, \mathbf{a})\in \mathcal{D}}$, during the test phase the function $f(\cdot)$ is evaluated over DOA $\hat{\bm{r}}$ and anthropometric features $\hat{\mathbf{a}}$ as
\begin{equation}\label{eq:hrir_estimate}
    \hat{h}(t,\hat{\bm{r}}, \hat{\mathbf{a}}) =  f_{\bm{\gamma^*}}\left(t, \hat{\bm{r}},\hat{\mathbf{a}} \right),
\end{equation}
it follows that a good predictor $f(\cdot)$ has to provide meaningful HRIR estimates for novel subjects ``unseen'' during the optimization and represented by features $\hat{\mathbf{a}}$.

In this work, we propose the use of a DDPM as HRIR estimator $f(\cdot)$ in \eqref{eq:problem_def}. 
The network is trained in order to generate HRIR samples based on the source DOA $\hat{\bm{r}}$ and the set of anthropometric features $\hat{\mathbf{a}}$ defining the target listener.

\section{Proposed Method}
\subsection{Conditioning data} \label{sec:conditining_data}
In order to generate the desired HRIR, the model is conditioned on DOA and ear measurement data. 
The spatial information is indicated via label $\bm{r}_l \in l=0\,\ldots,L$, corresponding to a finite set of $L$ possible DOAs.

For what concerns the anthropometric measurements, we use the same $N=27$ features indicated in\cite{algazi2001cipic}, of which $17$ belong to the head and torso, while $10$ are associated with each pinna.
Similarly to \cite{chun_deep_2017,wang2020global,Zhang_2023, chen2019autoencoding}, we normalize the anthropometric features $a_n, n=0,\ldots,N-1$ before feeding them to the network as
\begin{equation}\label{eq:sigmoid}
    \overline{a}_n = \frac{1}{1 + \mathrm{e}^{\left( -\frac{a_n - \mu_n}{\sigma_n}\right)}},
\end{equation}
where $\overline{a}_n$ is the normalized anthropometric feature, $\mu_n$ and $\sigma_n$ are the mean and standard deviation of the $n$th feature over all the subjects in the dataset.

\subsection{HRIR reconstruction via Diffusion model}
We adopt a conditional DDPM for the generation of the personalized HRIR. Specifically, we follow the original formulation proposed in~\cite{ho2020denoising}, where the training is split in a forward process, where noise is iteratively added to the input HRIR, and a backward process, where the HRIR is reconstructed from the noise. 

More formally, given the desired HRIR $\mathbf{h}$, the forward process consists of a fixed Markov chain that gradually adds gaussian noise to the input over $I$ iterations following a noise schedule defined by $\beta_i \in \mathbb{R}, i=0,\ldots,I-1$. At each iteration $i$ we can marginalize the forward process as follows
\begin{equation}
    q(\mathbf{h}_I|\mathbf{h}_0) = \mathcal{N}(\mathbf{h}_{i},\sqrt{\overline{\alpha}_i}\mathbf{h}_0, (1-\overline{\alpha})\mathbf{I}),
\end{equation}
where $\mathbf{h}_0$ is the starting HRIR without added noise,  $\overline{\alpha}_i=\prod_{s=1}^{i}(1-\beta_s)$ and $\mathbf{I}$ is the identity matrix.
Therefore, at each step index $i$, we can obtain $h_i$ as
\begin{equation}
    \mathbf{h}_i = \sqrt{\overline{\alpha}_i}\mathbf{h}_0 + \sqrt{1-\overline{\alpha}_i}\bm{\epsilon},
\label{eq:noise_step}
\end{equation}
where $\boldsymbol{\epsilon} \sim \mathcal{N}(\boldsymbol{\epsilon},\mathbf{0},\mathbf{I})$.

The backward process, instead, aims at reconstructing $\mathbf{h}_0$ from the noisy representations, but being $q(\mathbf{h}_i|\mathbf{h}_{i-1})$ an intractable problem, we resort to a neural network $f_\gamma$ that models the following equation
\begin{equation}
    P_{\bm{\gamma}} (\mathbf{h}_{0:I}) \triangleq P(\mathbf{h}_I) \prod_{i=1}^{I} P_{\bm{\gamma}} (\mathbf{h}_{i-1}|\mathbf{h_i}).
\end{equation}

Operatively, we train a conditional neural network $f_{\bm{\gamma}}(\mathbf{h}_i,i,\mathbf{r},\overline{\mathbf{a}})$ to estimate the noise  $\bm{\epsilon}$ added during \eqref{eq:noise_step}. The network parameters $\bm{\gamma}$ are retrieved by minimizing the following loss
\begin{equation}
    \mathbb{E}_{i,\bm{\epsilon}} = ||f_\gamma (\overline{\sqrt{\alpha}}_i\mathbf{h}_0 +\sqrt{1-\overline{\alpha}_i}\bm{\epsilon},i,\bm{r},\overline{\mathbf{a}}) ||_{2}^{2}.
\end{equation}

During the inference stage, the network is then fed with Gaussian noise and the conditioning data. The network predicts the noise that was added during the diffusion process, and this prediction is subtracted from the Gaussian noise to iteratively refine the signal. This denoising process reconstructs a new HRIR, providing a set of HRIRs that are generated without direct observation of the original data, but instead inferred from the model and conditioning data. 

A schematic representation of the training and inference procedures is shown in Fig.~\ref{fig:architecture}.

\subsection{Architecture}
The backward process is modeled using a modified U-Net architecture\cite{ronneberger_unet_2015} with symmetrical encoding and decoding paths adapted to work with time-domain signals. Each architecture block processes the input by (i) a 1D convolutional layer with ReLU activation and batch normalization, followed by (ii) a second convolutional layer with no activation. Both layers use a kernel size of 3 with padding to preserve the input size. In the downsampling blocks, a final convolution with kernel size 4 and stride 2 is applied to reduce the temporal resolution.

The encoder consists of five downsampling blocks, each with increasing channel sizes (4, 8, 16, 32, 64), reducing the temporal resolution at each stage. Conditioning information is incorporated through fully connected layers and added to the feature maps, achieving conditioning by concatenation. Skip connections are saved for the decoder, and self-attention layers with 4 attention heads are integrated after each downsampling block.

The decoder mirrors the encoder structure, employing transposed convolutions with the same parameters to progressively increase the temporal resolution. Skip connections and conditioning information are concatenated at each decoding output. The last 1D convolutional layer restores the feature maps to the original input dimensions.
It is worth noting that the network outputs two-channel signals representing the generated left and right HRIRs of the subject.
\section{Validation}
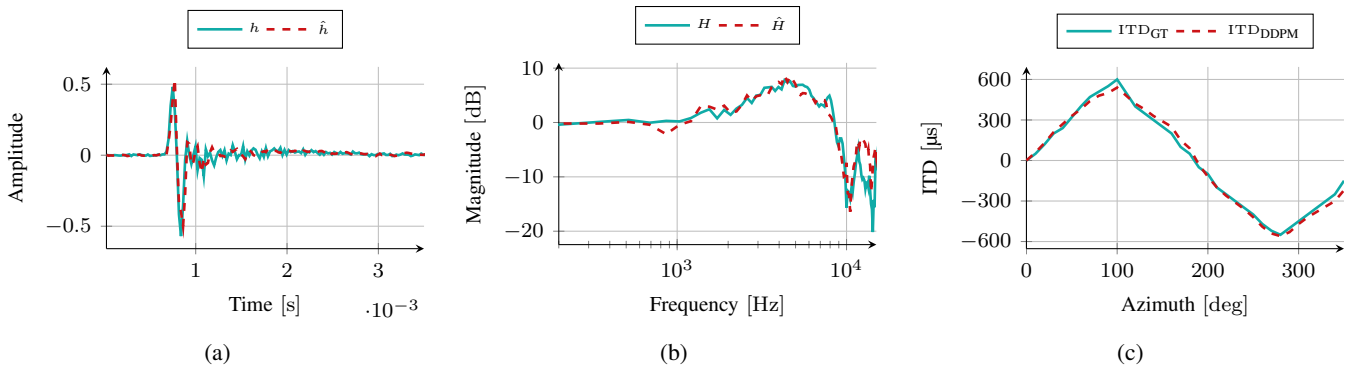
\begin{figure*}[htb]
    \centering
    \begin{subfigure}[b]{0.32\linewidth}
        \begin{tikzpicture}%
\begin{axis}[%
xmode=linear,%
xlabel={Time},%
x unit = \second,
ylabel={Amplitude},%
xmax=0.0035,
xtick={0,0.001,0.002,0.003,0.004,0.005},%
axis x line=bottom,%
axis y line=left,%
grid,%
height=4cm,%
enlarge y limits=0.08,%
width=\columnwidth,%
style={ font=\footnotesize},%
legend columns=2,%
legend style={at={(0.5,1.1)},anchor=south,font=\tiny},%
]
\addplot[draw={rgb,255:red,16;green,171;blue,167}, line width=1pt] table  {images/data/HRIR_gt.txt};%
\addlegendentry{$h$}%
\addplot[draw={rgb,255:red,204;green,19;blue,23}, dashed, line width=1pt] table  {images/data/HRIR_pred.txt};%
\addlegendentry{$\hat{h}$}%
\end{axis}%
\end{tikzpicture}%
        \caption*{(a)}
    \end{subfigure}
    \hspace{0.001\linewidth}
    \begin{subfigure}[b]{0.32\linewidth}
        \begin{tikzpicture}%
\begin{axis}[%
xmode=log,%
xlabel={Frequency},%
x unit = \hertz,
ylabel={Magnitude},%
y unit = \dB,%
xmin=200, xmax=15000,
axis x line=bottom,%
axis y line=left,%
grid,%
height=4cm,%
enlarge y limits=0.08,%
width=\columnwidth,%
style={ font=\footnotesize},%
legend columns=2,%
legend style={at={(0.5,1.1)},anchor=south,font=\tiny},%
]
\addplot[draw={rgb,255:red,16;green,171;blue,167}, line width=1pt] table  {images/data/HRTF_gt.txt};%
\addlegendentry{$H$}%
\addplot[draw={rgb,255:red,204;green,19;blue,23}, dashed, line width=1pt] table  {images/data/HRTF_pred.txt};%
\addlegendentry{$\hat{H}$}%
\end{axis}%
\end{tikzpicture}%
        \caption*{(b)}
    \end{subfigure}
    \hspace{0.001\linewidth}
    \begin{subfigure}[b]{0.32\linewidth}
        \begin{tikzpicture}%
\begin{axis}[%
xmode=linear,%
xlabel={Azimuth},%
x unit = \deg,
ylabel={ITD},%
y unit = \micro\second,%
ytick={-600,-300,0,300,600},
axis x line=bottom,%
axis y line=left,%
grid,%
height=4cm,%
enlarge y limits=0.08,%
width=\columnwidth,%
style={ font=\footnotesize},%
legend columns=2,%
legend style={at={(0.5,1.1)},anchor=south,font=\tiny},%
]
\addplot[draw={rgb,255:red,16;green,171;blue,167}, line width=1pt] table  {images/data/ITD_gt.txt};%
\addlegendentry{$\operatorname{ITD}_{\text{GT}}$}%
\addplot[draw={rgb,255:red,204;green,19;blue,23}, dashed, line width=1pt] table  {images/data/ITD_pred.txt};%
\addlegendentry{$\operatorname{ITD}_{\text{DDPM}}$}
\end{axis}%
\end{tikzpicture}%
        \caption*{(c)}
    \end{subfigure}
    \caption{(a) Subject 16 predicted $\hat{h}$ and ground truth $h$ for DOA $\bm{r}=(0,0)$, in (b) their respective HRTF and (c) ITD in the horizontal plane.}
    \label{fig:hrir}
\end{figure*}
\subsection{Setup}
We validated the performance of the proposed method using HUTUBS \cite{HUTUBS_dataset_2019}, a dataset of acoustically measured HRIRs of 93 subjects with complete anthropometric information, normalized as described in \ref{sec:conditining_data}. 
Likewise to \cite{chun_deep_2017, wang2020global}, we adopted the Leave-One-Out Cross-Validation (LOOCV) procedure in order to exploit the whole dataset information.

At the end of each LOOCV round, we generated the set of HRIRs relative to the test subject left out of the current training and validation dataset fold, selecting the weights of the model according to the best validation loss.

As far as the optimizer of the proposed U-Net architecture is concerned, we trained the network using Adam~\cite{kingma2014adam} with a learning rate equal to 0.001 and $20\%$ decay every 100 epochs. 

We trained the network for 1000 epochs for each subject (i.e. corresponding to each LOOCV round), and applied early stopping to finish the training if the validation loss did not improve for more than 200 consecutive epochs. Moreover, during the training and inference stages, we implemented $600$ noising and denoising steps with variance ranging from $1e^{\text{--}4}$ to $0.02$.

\subsection{Evaluation metrics}
We quantified the performance of our diffusion model-based HRTF personalization method in terms of Log-Spectral Distortion ($\LSD$) between the ground truth and the generated HRTF, being it a metric well suited to quantify the similarity between two magnitude spectra. As we trained in a LOOCV fashion, the $\LSD$ values considered the different training-validation rounds, that is, the results are averaged across all subjects and DOAs. The $\LSD$ is then computed as follows
\begin{equation}
    \LSD(H, \hat{H}) = \sqrt{\frac{1}{LK}\sum_l\sum_k\Bigg(20\log_{10} \Bigg| \frac{H(\bm{r}_l,k)}{\hat{H}(\bm{r}_l,k)}\Bigg|\Bigg)^2}
    ,
    \label{eq:lsd}
\end{equation}
where $\hat{H}$ and $H$ are the predicted and ground-truth HRTFs, for each one of the DOA $\bm{r}_l$; $k$ is the index of the frequency bins under analysis. 
Similarly to \cite{alvarez2024hrtf}, we evaluated $K = 44$ frequency bands between $0$ and $\SI{15}{\kilo\hertz}$. 

\subsection{HRIR personalization results}
In Fig.~\ref{fig:hrir} we show an example of one of the HRIR personalization outcomes obtained using the proposed method. 
In particular, it is possible to observe in Fig.~\ref{fig:hrir}(a) how the predicted HRIR approximates the ground truth one in terms of their onsets and amplitudes. 
This result is confirmed by Fig.~\ref{fig:hrir}(b), where we show the corresponding HRTFs, we can see that the two magnitude spectra follow the same trend with some deviations limited to frequencies over \SI{10}{\kilo\hertz}.
\begin{table}[tb]
\caption{Global $\LSD$}
\begin{center}
\begin{tabular}{|c|c|c|c|}
\hline
\cline{2-4} 
\textbf{} & \textbf{\textit{w/ SHT} \cite{wang2020global}}& \textbf{\textit{w/o SHT} \cite{wang2020global}}& \textbf{\textit{DDPM}} \\
\hline
$\LSD$ &\SI{4.74}{\dB}&\SI{6.06}{\dB} &\SI{5.1}{\dB}  \\
\hline
\end{tabular}
\label{tab:lsd}
\end{center}
\end{table}
Table~\ref{tab:lsd} presents the global $\LSD$ achieved by our method compared to the results from \cite{wang2020global}, which have been obtained through the same training approach.
Both methods rely solely on anthropometric features for personalization and are evaluated using HUTUBS after training in the LOOCV fashion. 
While the $\LSD$ reported in \cite{wang2020global} is marginally lower, our method deviates by only $\SI{0.36}{\dB}$ and outperforms \cite{wang2020global} without SHT.

\subsection{Further discussion on the personalization results}
Considering that our proposal generates personalized HRIR samples, we performed predictions of the subjects' left and right-side HRIRs to inspect the feasibility of predicting HRTF magnitudes and ITD within a single model. 
An example of ITD computed over predicted and ground truth HRIRs is shown in Fig.~\ref{fig:hrir} (c). The absolute error between the ground truth and predicted ITDs across all subjects and DOAs is $\SI{53.93}{\micro\second}$, which is in line with the $\SI{44.18}{\micro\second}$ error obtained in \cite{wang2022predictingglobalheadrelatedtransfer} and below 1 Just Noticeable Difference (JND) for various elevation angles $\phi$ according to \cite{johansson2022interaural}.

\begin{figure}[htb]
    \centering
    \begin{tikzpicture}%
\begin{axis}[%
xmode=log,%
xlabel={Frequency},%
x unit = \hertz,
ylabel={PBC mean},%
xmin=200, xmax=15000,
ymin=0, ymax=1,
axis x line=bottom,%
axis y line=left,%
grid,%
height=0.7*4cm,%
width=\columnwidth,%
style={ font=\footnotesize},%
legend columns=2,%
legend style={at={(0.5,1.1)},anchor=south,font=\tiny},%
]
\addplot[draw={rgb,255:red,16;green,171;blue,167}, line width=1pt] table  {images/data/PBC.txt};%
\end{axis}%
\end{tikzpicture}%
    \caption{Mean PBC computed across the ERB.}
    \label{fig:pbc}
\end{figure}
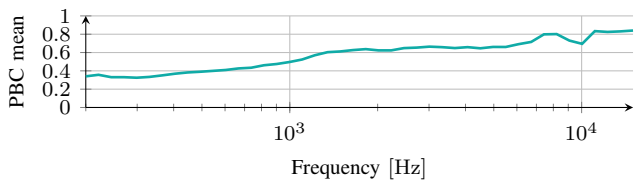

Through the Predicted Binaural Colouration (PBC) \cite{mckenzie_pbc_2022,marggraf2024hrtf} model included in the auditory modeling toolbox (AMT) \cite{Søndergaard_amt_2013}, we carried out an analysis from the perceptual point of view to understand how relevant the differences between our DDPM estimated HRIRs and the ground truth measurements are. 
Fig.~\ref{fig:pbc} depicts the mean colouration values associated with the Equivalent Rectangular Bandwidths (ERB) \cite{smith_bark_1999}, that is, representing the average perceptual difference in sones between the two compared spectra, as stated in \cite{mckenzie_pbc_2022}. 
We can notice that the colouration increases with the frequency value, implying higher values in the high-frequency range, where the auditory system is expected to be less sensitive \cite{william2007auditoryperception}. 
The HRTF in the high-frequency range might be influenced in a great extent by the ear shape, which is complex and may not be fully represented by the anthropometric measurements available.
\section{Conclusions}
In this paper, we proposed a denoising diffusion probabilistic model to tackle the HRTF personalization problem, leveraging this generative model by directly taking measured HRIRs and anthropometric measurements from a database to generate customized HRIR samples. 
Results demonstrate the ability of the proposed method in generating HRIRs that closely resemble the ground truth data. The objective evaluation indicated $\LSD$ values near the state-of-the-art benchmarks. Further analysis showed the feasibility of this proposal for predicting adequate ITD values and the PBC calculation pointed out that the reconstruction of the high-frequency range can be improved in future works. We encourage future research to explore the DDPMs to the HRTF personalization problem using alternatives to HRIR, such as HRTF magnitude or spherical harmonics transforms, as well as different representations of the subjects' anthropometric features as they can provide more relevant contributing factors to the HRTF. 
\ifCLASSOPTIONcaptionsoff
  \newpage
\fi
\balance
\bibliographystyle{IEEEtran}
\bibliography{biblio}
\end{document}